\documentclass{SCGE}

\begin{document}

\begin{picture}(0,0){\rm
\put(0,-39){\makebox[160truemm][l]{\bf {\sanhao\raisebox{2pt}{.}}
Article  {\sanhao\raisebox{1.5pt}{.}}}}}
\put(0,-52){\jiuwuhao {\textcolor[rgb]{0.5,0.5,0.5}{\sf %Nuclear Magnetic Moments
}}}%%(11ÔÂ×¢ÊÍ£ºµ÷\textcolor[rgb]{x,x,x}ÖеÄÊý×ÖxÔ½´óÔ½»Ò)
\end{picture}

\def\bm{\boldsymbol}

\def\dl{\displaystyle}
\def\du{\end{document}}
\def\pi{{\uppi}}

% The author doesn't need fill in it.
\Year{201?} %
\Month{March} %
\Vol{?} %  ¾íºÅ
\No{?} %  ÆÚºÅ
\BeginPage{0} % ÆðÒ³Âë
\EndPage{} %  Ö¹Ò³Âë
\AuthorMark{{\rm Zhou L.-Y. et al}}  %(11ÔÂ×¢ÊÍ£ºÒ³Ã¼ÉϵÄ×÷Õß)
\AuthorMarkCite{{\rm Zhou L.-Y., Li J., Cheng J., Sun Y.-S.}.} %(11ÔÂ×¢ÊÍ£ºcitationÖеÄ×÷Õß)
\DOI{10.1007/s11433-012-4648-2} % The author doesn't need fill in it.

% \title[short text for running head]{full title}{comments for title}
\title[Hyperbolic Structure and Stickiness Effect]{Hyperbolic Structure and Stickiness Effect: A case of a 2D Area-Preserving Twist Mapping}%±êÌâ

\author[*]{ZHOU Li-Yong$^{1,2}$, LI Jian$^{1,2}$, CHENG Jian$^3$ \& SUN Yi-Sui$^{1,2}$}{}
\footnote{Corresponding author (email: sunys@nju.edu.cn)}

\address[{\rm}]{1. School of Astronomy and Space Sciences, Nanjing University, Nanjing 210093, China \\
2. Key Laboratory of Modern Astronomy and Astrophysics in Ministry
of Education, Nanjing University, Nanjing 210093, China \\
3. Department of Mathematics, Nanjing University, Nanjing 210093,
China}

\maketitle \vspace{-3.5mm}{\footnotesize\begin{center} Received October 20, 2010; accepted March 3, 2011; published online February 8, 2012%ÊÕ¸åÈÕÆÚ
\end{center}}\vspace*{-5mm}

\begin{center}
\rule{16.5cm}{0.4pt}
\parbox{16.5cm}
{\begin{abstract} The stickiness effect suffered by chaotic orbits
diffusing in the phase space of a dynamical system is studied in
this paper. Previous works have shown that the hyperbolic structures
in the phase space play an essential role in causing the stickiness
effect. We present in this paper the relationship between the
stickiness effect and the geometric property of hyperbolic
structures. Using a two-dimensional area-preserving twist mapping as
the model, we develop the numerical algorithms for computing the
positions of the hyperbolic periodic orbits and for calculating the
angle between the stable and unstable manifolds of the hyperbolic
periodic orbit. We show how the stickiness effect and the orbital
diffusion speed are related to the angle.
%ÕªÒª
\end{abstract}}
\end{center}\vspace*{-0.6cm}

\begin{center}
\parbox{16.5cm}
{\bf\jiuhao stickiness effect, hyperbolic structure, stable and unstable manifolds}%¹Ø¼ü´Ê
\end{center}

\begin{center}
{}%·ÖÀàºÅ
\CITA    %%(11ÔÂ×¢ÊÍ£ºCitationÄÚÈÝ×Ô¶¯Éú³É)
%\Cit{}%%(11ÔÂ×¢ÊÍ£ºCitationÄÚÈÝÐèÊÖ¶¯Ìîд)
\end{center}

%%%%%%%%%%%%%%%%%%%%%%%%%%%%%%%%%%%%%%%%%%%%%%%%%%%%%%%%%%%%
\wuhao\vspace*{1.5mm}

\begin{multicols}{2}

%%%%%%%%%%%%%%%%%%%%%%%%%%%%%%%%%%%%%%%%%%%%%%%%%%%%%%%%%%%%
%% Text of article.
%%%%%%%%%%%%%%%%%%%%%%%%%%%%%%%%%%%%%%%%%%%%%%%%%%%%%%%%%%%%
%    Section headings
\renewcommand{\baselinestretch}{1.08} \baselineskip 12.2pt\parindent=10.8pt

\renewcommand{\thefootnote}

\section{Introduction}

In a $2n$-dimensional phase space of a dynamical system, given long
enough time, a chaotic orbit would visit any place in the connected
chaotic region, owning to the ergodicity property. Only those
regions surrounded by the $(2n-1)-$dimensional invariant surface
(curve) are forbidden to the chaotic orbit. During its diffusion in
the phase space, a chaotic orbit generally spends much longer time
wandering near the regular regions than elsewhere. This phenomenon
is called the ``stickiness effect'' [1, 2], and it causes the
anomalous chaotic transport [3--7]. The phase spaces of most
physically realistic systems consist of chaotic and regular
components, thus the stickiness effect is a generic phenomenon.
Especially, in those nearly integrable systems such as the Solar
System [8], the slow diffusion is a key point to understand the
stabilities of those systems. As a matter of fact, the diffusion
(transport) in the phase space is a very important issue in many
areas, with interests not only of the general theory of dynamical
system but also of direct applications in physics and astrophysics.
For example, this type of dynamics has implications in the tokamak
fusion [e.g. 9], turbulence [e.g. 10], plasma physics [e.g. 11],
semiconductor superlattices [e.g. 12], quantum chaos [e.g. 13],
galaxy dynamics [e.g. 14], etc.

The stickiness effect was first recognized in the close vicinity of
the KAM tori [1, 2], but it seemingly may arise from many other
structures in the phase space, including the island chains, the
cantori, the hyperbolic structures, and the asymptotic curves of the
unstable periodic orbits [see e.g. 2, 15--24]. We have shown in our
previous papers [25, 26] that the hyperbolic invariant sets play a
critical role in causing the stickiness effects. Using models of
both 2-dimensional (2D) mapping [25] and 3-dimensional (3D) mapping
[26], we have shown that a diffusing orbit is generally ``stuck'' by
the hyperbolic structures in the 2D or 3D phase space. There seems
to be an essential relation between the hyperbolic structures and
the stickiness effects.

Many studies on the stickiness effect focused on the diffusion
behaviors in the phase space, by illustrating what structures may
have stickiness effects [see e.g. 16, 20, 21, 23], or by describing
the rules of diffusion in different regimes [e.g. 8, 16, 17, 27]. As
for the geometric details of the structures causing the stickiness
effect, Contopoulos and Harsoula [23] calculated the sizes of gaps
in a cantorus being crossed by an orbit and relate the size of the
``maximum gap'' with the diffusion time.

Obviously the diffusion orbits have to pass through the ``gaps''
when crossing a cantorus. In a 2D phase space, the hyperbolic
periodic orbits are included in cantori and island chains. When an
orbit crosses these structures on its diffusion route, it approaches
from one side to the hyperbolic periodic point along the stable
manifold and then it may leave from the other side along the
unstable manifold. Therefore, the geometric features of the phase
space influence strongly the characteristic of diffusing
trajectories. In this paper, we try to explore the geometric
structures around the cantori, in particular, we study the angle
between the stable and unstable manifolds around the cantori.

To do this, it is necessary to know first the exact positions of the
cantori and the directions of the stable and unstable manifolds
mentioned above. But, since a cantorus is unstable, it's difficult
to locate its position precisely. Consequently, the directions of
the manifolds can hardly be determined accurately. Using a 2D
mapping model, we will introduce the methods for calculating such
positions and directions with high precision, then calculate the
angle between the directions of the stable and unstable manifolds,
and finally we will study the relation between the diffusion speed
and the sizes of the angles. We find that the diffusion speed
increases when the size of angles increases.

The mapping model used in this paper is obtained by a modification
to the so called ``standard mapping''. Contrary to the standard
mapping whose ``twist'' depends linearly on the action-variable, our
mapping model has a nonlinear twist. Such a nonlinearity in twist is
often seen in physics [28].

The paper is organized as follows. In Section 2, we introduce the
mapping model. We show in Section 3 how to locate the precise
positions of hyperbolic periodic orbits in the phase space and
present the method of calculating the angle between stable and
unstable manifolds. The characteristic angles of the hyperbolic
structures in different regions and the diffusion speeds of orbits
diffusing across these regions are compared in Section 4. And
finally we make the conclusions in Section 5.

\section{Mapping Model}\vspace*{-2mm}

\noindent Generally a mapping is much easier to handle than a set of
differential equations (e.g. a Hamiltonian system) as a dynamical
system. Thus we use a mapping model in this paper as usual.

\begin{figure}[H]
 \vspace{6.2 cm}
 \includegraphics{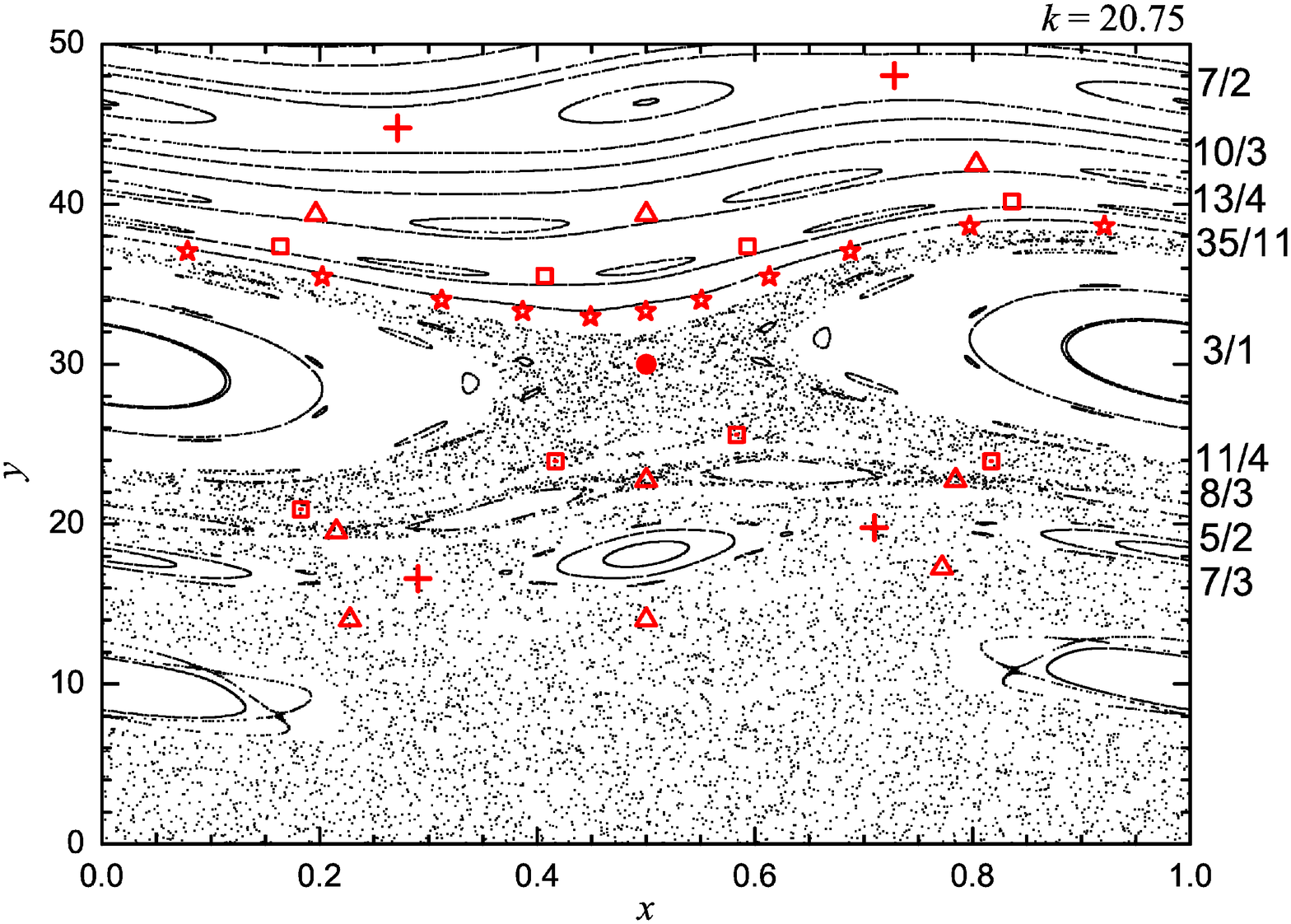}
 \caption{The phase space of the mapping (Eq.~(6)) with the parameter $k=20.75$.
 Some hyperbolic periodic orbits (fixed points) are
 located and plotted. To present a clean appearance,
 the period-two orbits are indicated by crosses, and the period-3, 4 orbits are indicated by
 triangles and squares, respectively. The 3/1 and the 35/11 periodic orbits are
 indicated by a solid dot and stars. The rotation number of the orbits are labeled on the
 right side.}
\label{minor}
\end{figure}

\subsection{The mapping}

\noindent We begin from the well-known standard mapping [29]:
 \begin{equation}
 \left\{
\begin{array}{l}
 \tilde{x}^\prime= \tilde{x}-\tilde{y}^\prime, \\
 \tilde{y}^\prime= \tilde{y}+\frac{k}{2\pi}\sin(2\pi \tilde{x}),
\end{array} \right.
 \end{equation}
which can be generated from a generating function
 \begin{equation}
\tilde{S}=\frac{1}{2}(\tilde{x}-\tilde{x}^\prime)^2+
\frac{k}{4\pi^2}\cos(2\pi \tilde{x}),
 \end{equation}
through equations
 \begin{equation}
\left\{
\begin{array}{l}
\tilde{y}=\tilde{S}_{\tilde{x}}=\frac{\partial \tilde{S}}{\partial
\tilde{x}}= \tilde{x}-\tilde{x}^\prime - \frac{k}{2\pi}\sin(2\pi
 \tilde{x}), \\
 \tilde{y}^\prime=-\tilde{S}_{\tilde{x}^\prime}=-\frac{\partial \tilde{S}}{\partial \tilde{x}^\prime}=
 \tilde{x}-\tilde{x}^\prime.
\end{array} \right.
\end{equation}
Here we modify the generating function by adding an extra high-order
term $\frac{1}{4}(x-x^\prime)^4$,
 \begin{equation}
S=\frac{1}{2}(x-x^\prime)^2+ \frac{1}{4}(x-x^\prime)^4+
\frac{k}{4\pi^2}\cos(2\pi x),
 \end{equation}
thus the corresponding area-preserving mapping is given by
 \begin{equation}
 \left\{
\begin{array}{l}
 y=S_x=(x-x^\prime)+(x-x^\prime)^3-\frac{k}{2\pi}\sin(2\pi x),\\
 y^\prime=-S_{x^\prime}=(x-x^\prime)+(x-x^\prime)^3.
\end{array} \right.
 \end{equation}
\noindent Explicitly, the mapping reads
 \begin{equation}
 \left\{
\begin{array}{l}
 x^\prime=x- \sqrt[3]{\frac{y^\prime}{2}+\sqrt{\frac{y^{\prime 2}}{4}
 +\frac{1}{27}}} - \sqrt[3]{\frac{y^\prime}{2}-\sqrt{\frac{y^{\prime 2}}{4}
 +\frac{1}{27}}} \hspace{0.5 cm} {\rm mod}(1), \\
 y^\prime=y+\frac{k}{2\pi}\sin(2\pi x).
\end{array} \right.
 \end{equation}
\noindent This mapping is defined on the cylinder $0 \leq x \leq 1,
-\infty < y < +\infty$, and $k$ is the only one perturbation
parameter. As in the standard mapping, the $y$ in this mapping is
the action variable, and the $x$ is the angle variable that can be
performed the modulus over 1. As usual, in this paper, by
``diffusion'' we mean the drifting of the action variable $y$ in the
phase space. In the study by Cheng and Sun [30], it was shown that
for any given $k$, this mapping behaves chaotically around $|y|<M$
(where $M>0$), and the KAM curves exist at large $|y|$. When $k$ is
small, the phase space of the mapping is mainly full of (horizontal)
invariant tori. As $k$ increases, the KAM tori break, first at low
$|y|$ value, and chaos sets in. The area occupied mainly by chaotic
orbits extends outward from the region around $y=0$. We show in
Fig.~\ref{minor} the phase space when $k=20.75$. Note that the
negative $y$ half of the phase space is symmetric to the positive
half that has been shown in Fig.~\ref{minor}.

\begin{figure}[H]
 \vspace{5.2 cm}
 \includegraphics{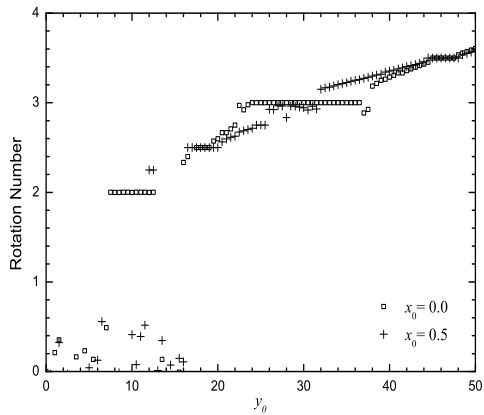}
 \caption{The rotation number of the mapping ($k=20.75$). Open squares
 show the change of rotation number along the vertical line $x=0.0$, while the crosses
 are for the line $x=0.5$.}
 \label{rotn}
\end{figure}

The mapping is an area-preserving monotone twist mapping, thus along
a vertical line in the phase space the rotation number increases
monotonically. The rotation numbers along two vertical lines are
computed and shown in Fig.~\ref{rotn}. Clearly, the scattering
distribution in the lower left corner of Fig.~\ref{rotn} is a
reflection of the chaotic motion at low $y$ value in the phase
space. The ``plateaus'' in both cases arise from the fact that the
vertical lines cross stable islands, on which the rotation numbers
are constants.

\subsection{The orbital diffusion in the phase space}

Any chaotic orbit in the connected chaotic sea of the phase space
will wander ergodically as time tends to infinity. But, it is known
that an orbit spends much longer time in some specific regions than
elsewhere, e.g., the close vicinity around an embedded island, the
region where a KAM torus had newly broken, the region occupied by
hyperbolic structures, etc. This is called the stickiness effect, as
we mentioned above.

\begin{figure}[H]
 \vspace{5.2 cm}
 \includegraphics{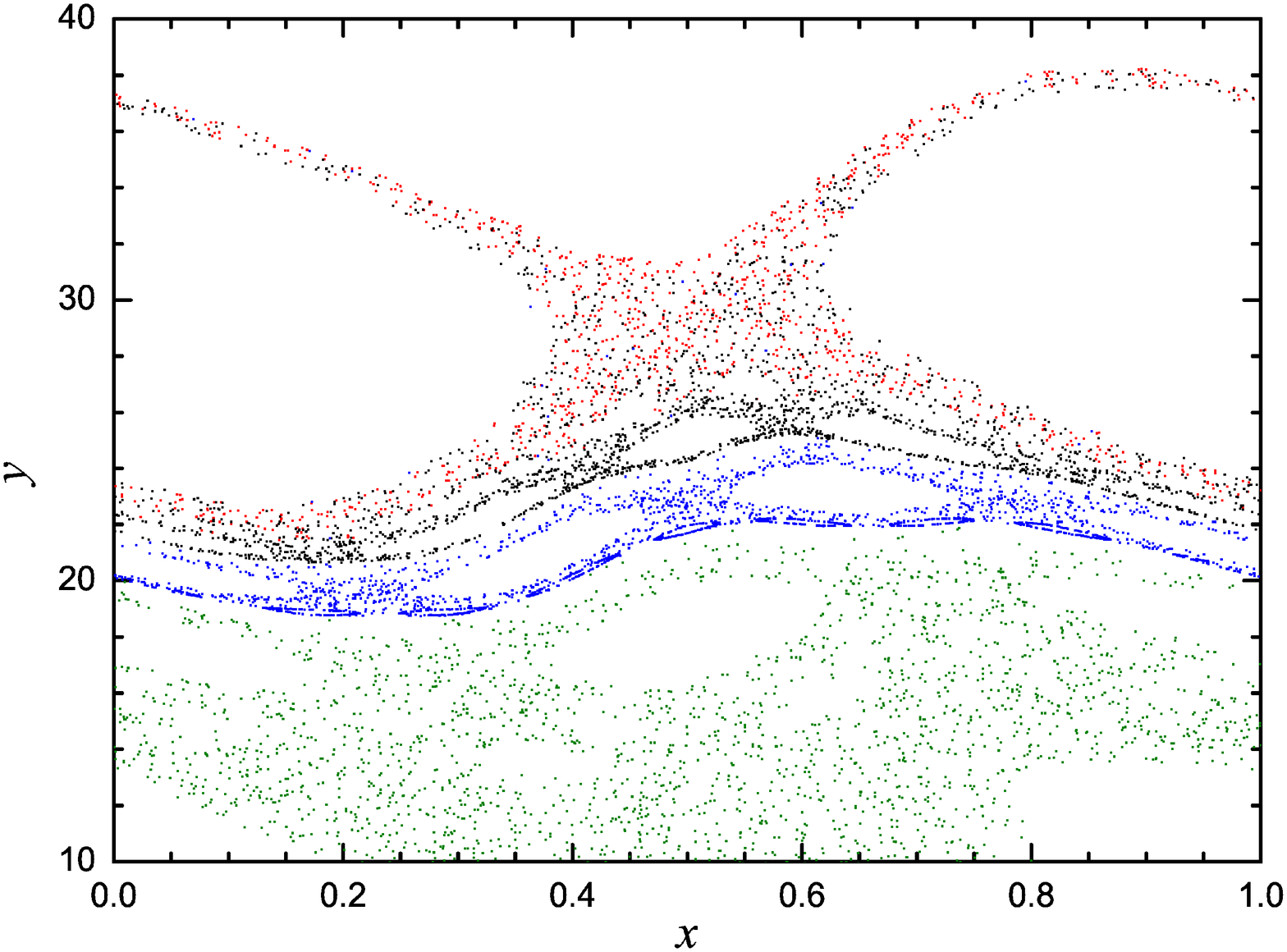}
 \caption{The diffusion process of an arbitrary orbit starting from
 $(x_0,y_0)=(0.49,29.0)$. The red, black and blue points represent its trajectories during the
 time (iteration number $n$) intervals of $(1,9\times 10^4), (9\times 10^4, 4.5\times
 10^7)$ and $(4.5 \times 10^7, 7 \times 10^7)$. After $\sim 6\times 10^9$ iterations,
 it finally crosses the compact partial barrier (see text), which makes the bottom boundary of blue
 points, and enters the open chaotic sea around $y=0$ (green dots). }
 \label{difr}
\end{figure}

To show a typical orbital diffusion process and the stickiness
effect, we arbitrarily select an initial point
$(x_0,y_0)=(0.49,29.0)$ not far away from the regular region in the
upper part of Fig.~\ref{minor}, and follow its evolution. The
diffusion of this orbit is shown in Fig.~\ref{difr}.

On its route of downward diffusion, the orbit meets some structures
that play the role of obstacles. As shown in Fig.~\ref{difr}, the
orbit wanders around the 3/1 fixed point and the corresponding
island (see Fig.~\ref{minor} to find the positions of island-chains
of different rotation numbers) at beginning (red points trajectory).
This island does not show a significant stickiness effect, but the
orbit spends much longer time in next stage around the vicinity of
the 11/4 island-chain (black dots in Fig.~\ref{difr}). After that,
it was obstructed by another barrier, which prevents the orbit from
further diffusing downward, and only after pretty long time, the
orbit can reach the 8/3 island chain (blue dots). Then, the
diffusion was strongly obstructed by a partial barrier below the 8/3
island chain. Only after $\sim 6\times 10^9$ iterations, the orbit
can cross this ``compact'' barrier and enter finally the open
chaotic sea around $y=0$ (green dots). In fact, the rotation number
around this compact partial barrier is around 2.6 (see
Fig.~\ref{rotn}). We suspect that this difficult-to-cross barrier is
the remnant of the last and most robust KAM torus that has the
``golden rate'' rotation number. And this will be verified in next
section.

One thing we would like to address here is that the orbit may cross
an obstacle in both forward and backward directions, that is, an
orbit may cross the same partial barrier several times back and
forth. After leaving a region, it may re-enter the region again some
time later. That's why we see the black dots and red dots are mixed
around the 3/1 island in Fig.~\ref{difr}.

\begin{figure}[H]
 \vspace{3.7 cm}
 \includegraphics{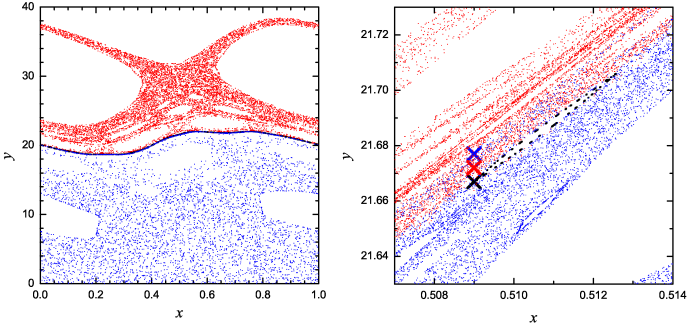}
 \caption{The most sticky region. Three initial points (crosses) and the
 corresponding trajectories (dots) are shown in black, red and blue.}
\label{tori}
\end{figure}

Typically, an orbit starting above the lowest and difficult-to-cross
barrier mentioned above would spend time as long as several $10^9$
iterations before crossing it. In attempt to show some details of
this barrier, we plot in Fig.~\ref{tori} three orbits starting close
to it, from initial points $P_1=(0.509,21.667), P_2=(0.509,21.672)$
and $P_3=(0.509,21.677)$ respectively. These initial points are
indicated by three crosses in Fig.~\ref{tori}b. From this magnified
picture, we realize that $P_3$ (black) is on an island-chain with a
high period number. The orbit initialized at $P_2$ (red) finally
diffuses upward after several $10^9$ iterations, while the one from
$P_1$ (blue) diffuses downward. There is a mixed region where we can
find both red and blue dots, indicating clearly the existence of the
difficult-to-cross partial barrier.

\section{Angle between stable and unstable manifolds}

The diffusion of an orbit is through the hyperbolic structures
between the islands in an island-chain, which is in fact the relic
of the KAM tori if the rotation number in this position does not
satisfy the Diophantine condition [31]. On the other hand, if the
rotation number is irrational enough, there would be a cantorus left
after the breaking of the KAM torus. And an orbit's diffusion is
through the gaps in the cantorus. But the cantorus itself is
composed of hyperbolic points and in the gaps exist the hyperbolic
structures.
%In this sense, all the stickiness
%effects are caused by hyperbolic structures as we argued in our
%previous papers \citep{sun05,sun09}.

In this paper, we will describe the geometric property of the
hyperbolic structures and find out how this property affects the
diffusion speed. We will show that the diffusion speed is determined
by the angle between the stable and unstable manifolds of the
hyperbolic structure. In this section, we introduce the algorithm
for calculating the angle between the stable and unstable manifolds
of a hyperbolic structure.

\subsection{The locations of hyperbolic periodic points}

Usually a recursive algorithm is employed to compute the locations
of periodic orbits. But it does not work for the hyperbolic ones
because these orbits are unstable. Fortunately, the hyperbolic
periodic orbits we are looking for in this model are of minimal
action, therefore the gradient algorithm could be used to find these
orbits. The existence of these minimal periodic orbits is guaranteed
by Aubry-Mather theory [32].

Suppose $x_0, x_1, \cdots, x_{p-1}, x_p$ are the variables in the
configuration space of a minimal periodic orbit of rotation number
$q/p$, with $q$ and $p$ being coprime integers. Obviously,
$x_p=x_0+q$. The action functional $f$ corresponding to this
periodic orbit is:
 \begin{equation}
f(x_0, x_1, \cdots, x_{p-1})= S(x_0,x_1)+ S(x_1,x_2)+ \cdots+
S(x_{p-1},x_p),
 \end{equation}
\noindent where $S$ is the generating function (Eq.~(4)). Then the
minimal periodic configuration $(x_0, x_1, \cdots, x_{p-1}, x_p)$ is
the minimum of $f$. Therefore, a periodic orbit of rotation number
$q/p$ can be found by locating the minimum of the above mentioned
functional. To find the minimum, we may use the gradient algorithm.

Consider the following differential equation:
 \begin{equation}
\dot{\Xi}=-\nabla f(\Xi)
 \end{equation}
where $\Xi=(\xi_0,\xi_1,\cdots,\xi_{p-1})^T$ is a column vector and
the dot over it indicates time derivative. A trajectory defined by
this differential equation converges to a minimum of the functional
$f$ as $t \rightarrow \infty$. Thus, a minimal periodic orbit can be
obtained through integrating this differential equation.

Specifically in our model, since
 \begin{equation}
 S_{x_k}(x_{k-1},x_k)=-y_k, \, S_{x_k}(x_k,x_{k+1})=y_k, \, k=1,2,\dots,p.
 \end{equation}
\noindent where the subscript denotes the partial derivative over
the corresponding variable, and
 \begin{equation}
 f_{x_k}=S_{x_k}(x_{k-1},x_k)+S_{x_k}(x_k,x_{k+1})=-y_k+y_k=0,
 \end{equation}
\noindent the corresponding differential equations reads:
\end{multicols}
 \begin{equation}
 \left\{
\begin{array}{l}
 \dot{\xi}_0=-(\xi_0-\xi_1)-(\xi_0-\xi_1)^3+\frac{k}{2\pi}\sin(2\pi \xi_0) + [\xi_{p-1}-(\xi_0+q)]+ [\xi_{p-1}-(\xi_0+q)]^3,\\
 \dot{\xi}_1=-(\xi_1-\xi_2)- (\xi_1-\xi_2)^3 + \frac{k}{2\pi}\sin(2\pi \xi_1)+(\xi_0-\xi_1)+(\xi_0-\xi_1)^3,\\
 \cdots, \cdots \\
 \dot{\xi}_n= - (\xi_n-\xi_{n+1})- (\xi_n-\xi_{n+1})^3 + \frac{k}{2\pi}\sin(2\pi \xi_n)+(\xi_{n-1}-\xi_n)+(\xi_{n-1}-\xi_n)^3,\\
 \cdots, \cdots \\
 \dot{\xi}_{p-2}= - (\xi_{p-2}-\xi_{p-1})- (\xi_{p-2}-\xi_{p-1})^3 + \frac{k}{2\pi}\sin(2\pi \xi_{p-2})+(\xi_{p-3}-\xi_{p-2})+(\xi_{p-3}-\xi_{p-2})^3,\\
 \dot{\xi}_{p-1}=-[\xi_{p-1}-(\xi_0+q)]-[\xi_{p-1}-(\xi_0+q)]^3+\frac{k}{2\pi}\sin(2\pi \xi_{p-1})
 + (\xi_{p-2}-\xi_{p-1})+ (\xi_{p-2}-\xi_{p-1})^3.
\end{array} \right.
 \end{equation}
\begin{multicols}{2}
The solution of these differential equations can be computed
numerically. We adopt the 7th order Runge-Kutta-Fehlberg algorithm
(RKF7(8)) to integrate the differential equations to time $10^4$,
unless the solution reaches the convergent value within a given
error when we regard the solution attains the required accuracy. In
this paper, we set the controlling error to be $10^{-14}$, that is,
if the difference between the successive values obtained from the
integrator is smaller than $10^{-14}$, we regard the convergent
solutions have been achieved.

\begin{figure}[H]
 \vspace{4.5 cm}
 \includegraphics{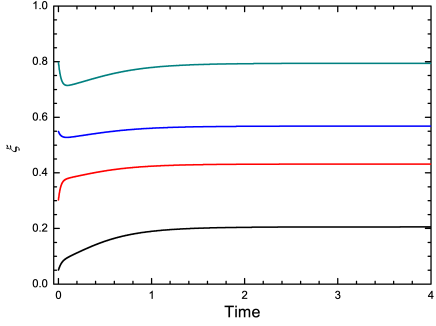}
 \caption{The convergence of the solution to the periodic orbit of rotation number 9/4.}
 \label{converg}
\end{figure}

In fact, the solution converges quite quickly. We show in
Fig.~\ref{converg} an example of the convergence of the trajectory
to the minimal periodic orbit of rotation number 9/4. We start from
$(\xi_0=0.05, \xi_1=0.30, \xi_2=0.55, \xi_3=0.80)$, and the solution
converges to the final value very soon, at integrating time $t\sim
11$.

With the $x$ values of the minimal periodic orbit, the corresponding
$y$ values can be computed using the generating function, via
Eq.\,(5). And finally we get the positions of the minimal periodic
orbit, i.e. the hyperbolic periodic orbits in the phase space. We
have over-plotted some of these calculated points on the phase space
as shown in Fig.~\ref{minor}.

\subsection{Characteristic angle of the hyperbolic structure}

Knowing the positions of hyperbolic periodic orbits, we now turn to
their properties. One of the most important characters of the
hyperbolic structure is the angle making by directions of the stable
and unstable manifolds, which will be called ``characteristic
angle'' for short hereafter. The directions of stable and unstable
manifolds are determined by two special tangent vector fields along
the periodic orbit, while the tangent vector fields can be obtained
by a limit processes. Along an orbit, the projection of the tangent
vector field onto the horizontal direction gives a Jacobi field,
which can be easily calculated by using the generating function of
the mapping.

We start our computations with the definition of Jocobi fields:
Assume a hyperbolic periodic orbit $\{x_n\}_{n=-\infty}^\infty$ of
rotation number $q/p$ has been found, and it satisfies
\[
 x_0, x_1, \cdots, x_{p-1} \quad  {\rm and} \quad x_{n+p}=x_n+q \quad (n=0, \pm 1, \pm 2,
 \cdots).
\]
Then,  $\{\xi_n\}_{n=-\infty}^\infty$ is said to be a ``Jacobi
field'' along the orbit $\{x_n\}_{n=-\infty}^\infty$ if
\end{multicols}
 \begin{equation}
 S_{12}(x_{n-1},x_{n})\xi_{n-1} + \left[S_{22}(x_{n-1},x_{n}) +
 S_{11}(x_{n},x_{n+1})\right]\xi_{n}
 +S_{12}(x_{n},x_{n+1})\xi_{n+1}= 0,
 \end{equation}
where $S$ is the generating function Eq.\,(4) and the subscript `1'
and `2' indicate the partial derivatives with respect to the first
and second variable of the function. Actually, the above equations
can be written in a matrix form. Set
 \begin{equation}
A_n= S_{22}(x_{n-1}, x_n)+ S_{11}(x_{n}, x_{n+1}), \quad
B_n=S_{12}(x_n,x_{n+1}), \quad  n=0, \pm 1, \pm 2, \cdots
 \end{equation}
\begin{multicols}{2}
Then we may write the equations as:
 \begin{equation}
 \left(\begin{array}{ccccccc}
 \ddots & \ddots & \ddots & \ddots & \ddots &  &  \\
 \cdots & 0 & B_{n-1} & A_n & B_n & 0 & \cdots \\
   &  & \ddots & \ddots & \ddots & \ddots & \ddots
 \end{array}\right)
 \left(\begin{array}{c}
  \vdots  \\
  \xi_{n-1} \\
  \xi_{n} \\
  \xi_{n+1} \\
  \vdots
 \end{array}\right)
 =0
 \end{equation}
Any finite piece of this triangular matrix is positive definite if
the corresponding configuration $\{x_n\}$ is minimal. A Jacobi field
along a minimal configuration induces an orbit of the corresponding
tangent map [33].

To compute the Jacobi fields along a minimal periodic orbit
$\{x_n\}$, which correspond to the stable or unstable directions, we
can get the approximation of the real Jacobi fields by solving the
following linear equations:
 \begin{equation}
 \left(\begin{array}{ccccc}
 A_1 & B_1 & 0 & \cdots & 0  \\
 B_1 & A_2 & B_2  & \ddots & \vdots \\
 \vdots &  \ddots & \ddots & \ddots & \vdots \\
 \vdots & \ddots & B_{n-2} & A_{n-1} & B_{n-1} \\
  0  & \cdots & 0 & B_{n-1} & A_n
 \end{array}\right)
 \left(\begin{array}{c}
  \xi_1 \\
  \xi_2 \\
  \vdots \\
  \xi_{n-1} \\
  \xi_n \\
 \end{array}\right)
 =
 \left(\begin{array}{c}
  -B_0 \\
  0 \\
  \vdots \\
  0 \\
  0 \\
 \end{array}\right)
 \end{equation}

As $n\rightarrow\infty$, we get the real Jacobi field. Note that
$\xi_1$ in above equations is monotonically increasing as $n$ goes
to infinity. In fact, the limit value of $\xi_1$ is the projection
on $x-$axis of the stable direction at point $(x_1,y_1)$. Since
(from Eq.\,(5))
\[
y_0=S_1(x_0,x_1),
\]
the projection on $y-$axis of the stable direction at point
$(x_0,y_0)$ can be computed from
\begin{equation}
\eta_0=S_{11}(x_0,x_1)\xi_0+S_{12}(x_0,x_1)\xi_1,
\end{equation}
with $\xi_0=1$. Finally the stable direction at $(x_0,y_0)$ is
defined by $\vec{\rm r}_s=(1,\eta_0)$. Note that this Jacobi field
could be used to calculate the Lyapunov exponent $\lambda$ of the
orbit, where $\lambda=- \lim_{n\to\infty} \frac{1}{n}\ln{\frac{\xi_n
}{\xi_0}}$.
% For details of the computation, refer to \citet{chen04}.

\begin{figure}[H]
 \vspace{4. cm}
 \includegraphics{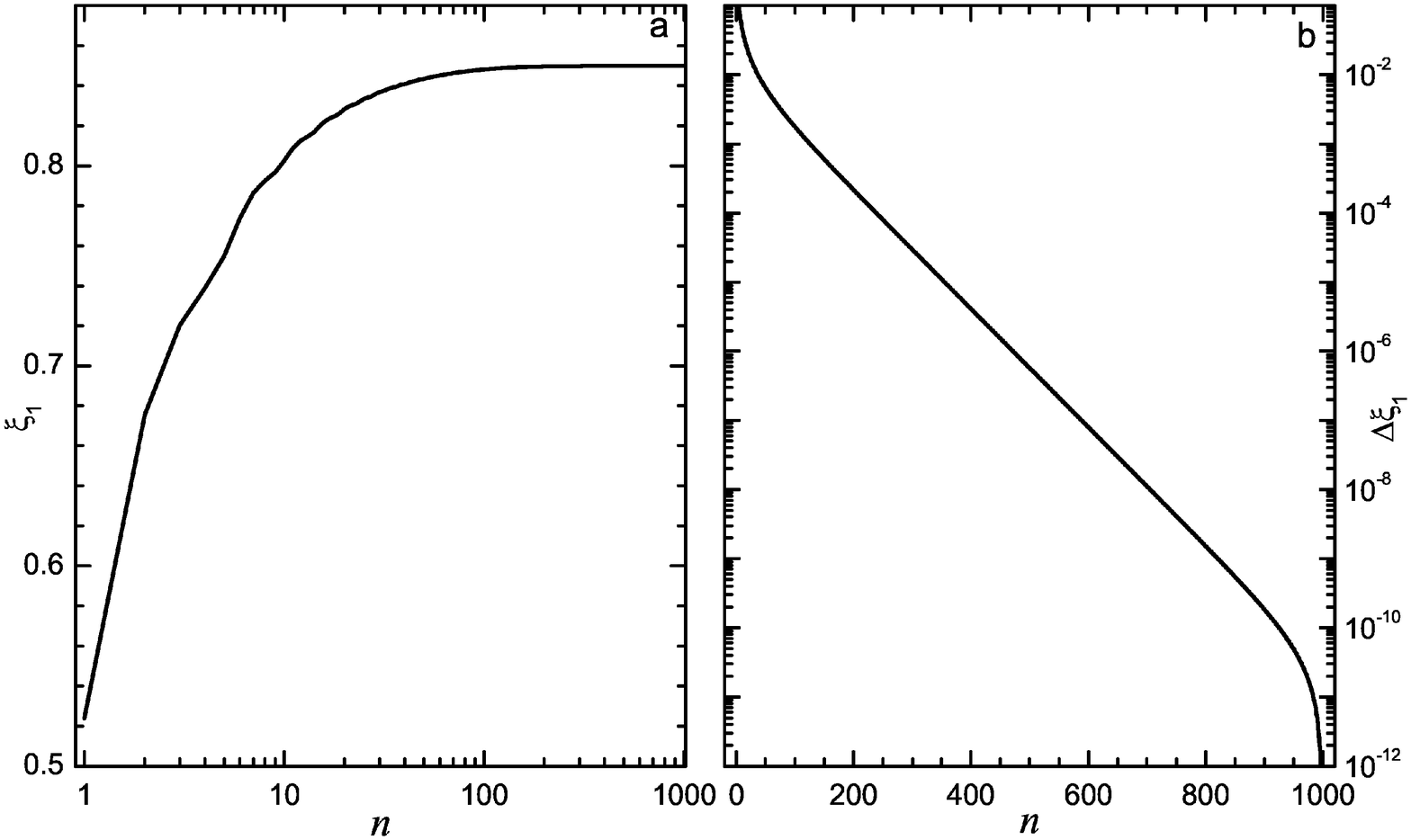}
 \caption{The convergence of $\xi_1$ in Eq.~(15). This is the case for the minimal periodic
 orbit of rotation number $9/4$. (a) $\xi_1$ versus recursive number $n$;
 (b) The error versus $n$, where the error is defined as the difference
 $\Delta\xi_1= \xi_1(n)-\xi_1(1000)$.}
 \label{xicon}
\end{figure}

We calculate $\xi_1(n)$ using a recursive algorithm. An example for
the minimal periodic orbit of rotation number $9/4$ is shown in
Fig.~\ref{xicon}. The convergence of $\xi_1$ with the increasing
recursive number $n$ can be seen clearly.
%Therefore, we use $\xi_1(n)$ for large $n$ as the limit value.

In a similar way, reversing the direction of calculation, we obtain
$\xi_{-1}$, which is the projection of unstable direction on
$x-$axis at point $(x_{-1},y_{-1})$. Considering
\[
y_0=-S_2(x_{-1},x_0),
\]
we compute the $y-$axis projection of the unstable direction at
point $(x_0,y_0)$ through:
\begin{equation}
\eta^\prime_0 = -S_{12}(x_{-1},x_0)\xi_{-1}-S_{22}(x_{-1},x_0)\xi_0
\end{equation}
with $\xi_0=1$. And the unstable direction at point $(x_0,y_0)$ is
$\vec{\rm r}_u=(1,\eta^\prime_0)$.

Finally, the angle between the stable and unstable directions, i.e.
the characteristic angle of the hyperbolic structure, can be
calculated easily
\begin{equation}
\alpha = \arccos\left(\frac{\vec{\rm r}_s\cdot\vec{\rm
r}_u}{|\vec{\rm r}_s|\,|\vec{\rm r}_u|}\right).
\end{equation}

As examples of the above algorithm, we calculated the positions of a
series of hyperbolic periodic orbits and their characteristic
angles. Generally, the more irrational a rotation number is, the
more robust the corresponding KAM torus is, and the stronger the
stickiness effect is around the relics of the KAM torus after its
breaking. People often write the rotation number in the form of
continued fraction, on account of the merit of its direct relation
to the ``irrationality''. A continued fraction
\[
a_0+\frac{1}{a_1 +\frac{1}{a_2+\frac{1}{a_3+\cdots}}}
\]
can be denoted by $[a_0;a_1,a_2,a_3,\cdots]$, where $a_0$ is the
integer part of the number and $a_i$ are the positive integers. Any
rational number can be expressed as a truncated continued fraction
$[a_0;a_1,a_2,a_3,\cdots,a_n]$. And the integer $n$ will be called
the ``order'' of the continued fraction in this paper. Note an
$n$\,th order continued fraction ended with $a_n=2$ is the same as
an $(n+1)$\,th order one ended with $a_n=1, a_{n+1}=1$.

\begin{table*}[t]
\caption{Angles between the stable and unstable manifolds of
hyperbolic periodic orbits. The first 3 columns are the rotation
numbers given in different forms, the forth column is the order of
the continued fraction, and the last column is the angle (in
degrees).}
 \center{\begin{tabular}{|l|l|l|c|r|}
 \hline
 Rotation Number & Fraction & Continued Fraction & Order & Angle \\
 \hline
 2.5  & 5/2 & [2;2] & 1 & 169.573 \\
 2.6666$\cdots$ & 8/3 & [2;1,2] & 2 & 116.235 \\
 2.6 & 13/5 & [2;1,1,2] & 3 & 9.034 \\
 2.625 & 21/8 & [2;1,1,1,2] & 4 & 2.094 \\
 2.6153$\cdots$ & 34/13 & [2;1,1,1,1,2] & 5 & 1.042 \\
 2.6190$\cdots$ & 55/21 & [2;1,1,1,1,1,2] & 6 & 1.657 \\
 2.6176$\cdots$ & 89/34 & [2;1,1,1,1,1,1,2] & 7 & 0.742 \\
 2.6186$\cdots$ & 144/55& [2;1,1,1,1,1,1,1,2] & 8 & 0.583 \\
 \hline
 \end{tabular}}
 \label{examp}
\end{table*}

Several hyperbolic periodic orbits are calculated and the results
are listed in Table~\ref{examp}. An obvious tendency derived from
Table~\ref{examp} is that, the bigger period number (the integer $p$
of the rotation number $q/p$) leads to the smaller characteristic
angle. As the rotation number is getting more and more irrational
the angle becomes smaller and smaller. It is well-known nowadays
that the last KAM tori are those with the ``noble'' rotation numbers
[34], i.e. the continued fractions that have $a_k=1$ for all $k$
above a certain number $N$. And the torus with the ``golden rate''
rotation number $(\sqrt{5}-1)/2 =
\left[0;1,1,1,\cdots,1,\cdots\right]$ is surely the most robust one.
In our model, we see in Table~\ref{examp} that the angle becomes
smaller when the rotation number approaches this noble value. And
also, we notice that the ``difficult-to-cross partial barrier'' and
the most sticky region we mentioned in \S~2.2 are just the region
corresponding to this noble rotation number.

\section{Stickiness effect and characteristic angle of hyperbolic structure}

When the trajectory of an orbit is plotted on the phase space, we
can see the trajectory points accumulate in some specific areas,
indicating the existence of some structures causing stickiness
effect. In this way we may obtain a direct impression of the sticky
region, just as we have shown in Fig.~\ref{difr}. Considering a
bunch of numerous orbits in the phase space, they may leave a
certain region as time passes by. The survival probability of orbits
in the definite region decays with time slowly. The slow decay is
also an indicator of the stickiness effect [16, 17, 25]. Following
these techniques, in this section we will show the diffusion routes
of some typical orbits and find out where they are ``stuck''. We
will check the structures causing the stickiness effect in detail,
and figure out the relation between the diffusion speed (or
inversely, the stickiness effect) and the characteristic angle of
hyperbolic structures in the sticky region.

\subsection{Typical diffusion route}

Due to the chaotic character, a chaotic orbit in its diffusion route
may diffuse toward any direction at a moment. % in the phase space.
Since the upper part of the phase space is not open to a chaotic
orbit but barred by KAM tori as we have seen in Fig.~\ref{minor}, an
orbit will diffuse downward overall. To monitor the diffusion, we
check the minimal $y$ value that an orbit attains on its trajectory
in the phase space. During the diffusion, each time when the $y$
value reaches a new minimum $y_{\rm min}$, we record the time $t$.
In such $(t,y_{\rm min})$ records, the ``back and forth'' motion in
the diffusion is abandoned, and the ``one-way'' diffusion is
emphasized.
%We may now focus on the stickiness effect that an orbit meets during
%the diffusion.

\begin{figure}[H]
 \vspace{6.2 cm}
 \includegraphics{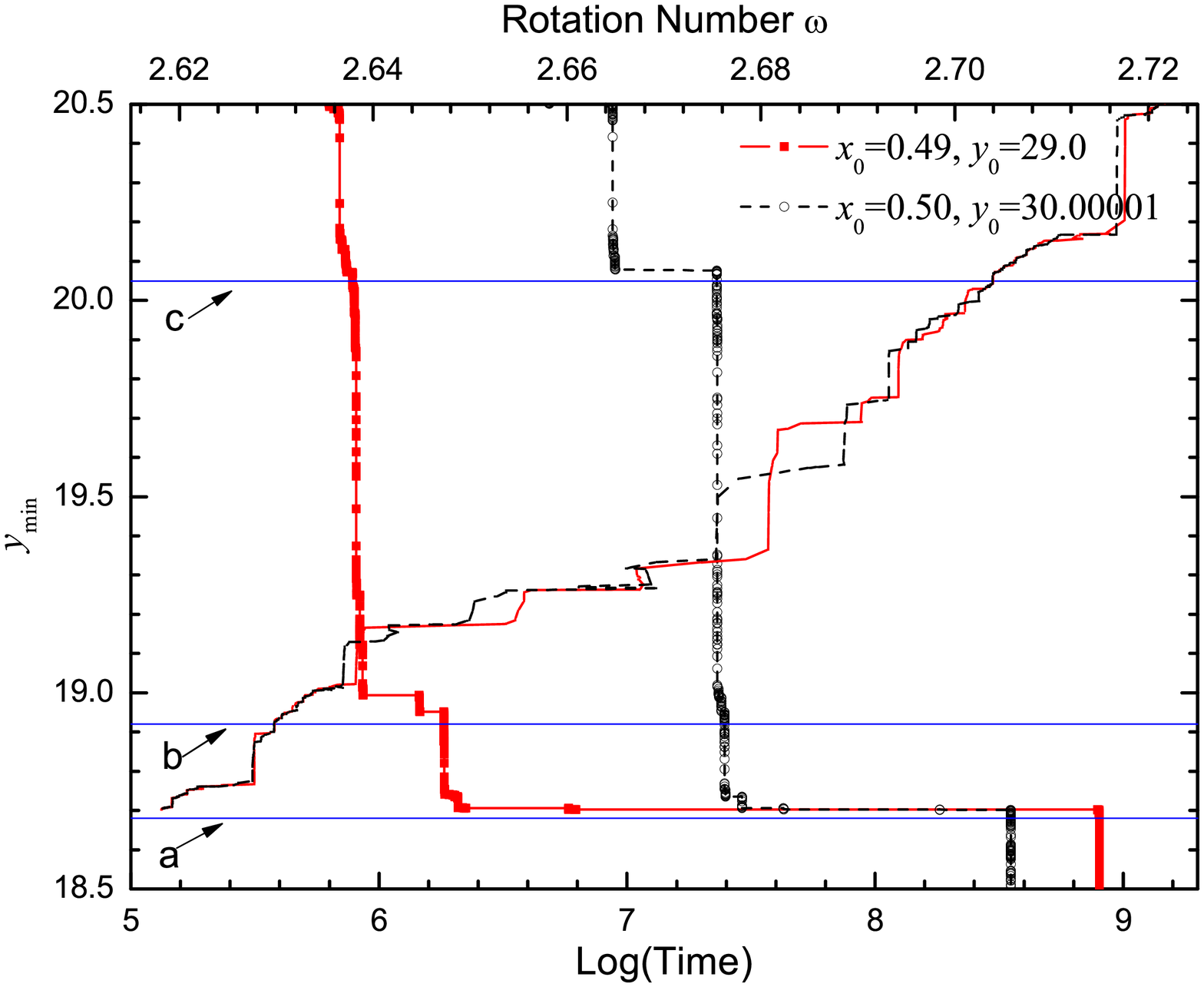}
 \caption{The variation of minimal $y$ of an orbit with respect to time (curves from
 upper left to lower right). The red solid squares are for the orbit
 initialized at $(0.49,29.0)$ and the black open circles for
 $(0.50,30.00001)$. The rotation numbers (indicated on the top $x$-axis) of orbits around
 the corresponding local lowest position are plotted (two curves
 crossing the panel from the lower left to upper right). Three regions that
 will be analyzed below are indicated by horizontal lines at
 $y_{\rm min}=20.05, 18.92$ and $18.68$ (see text for explanations). }
 \label{ymin}
\end{figure}

Arbitrarily, two initial points $(x_0, y_0)=(0.49, 29.0)$ and
$(0.50, 30.00001)$ are selected and their diffusions are followed.
The temporal evolutions of the minimums of the $y$ value are shown
in Fig.~\ref{ymin}. Note only part of the orbits, when the minimums
are smaller than $y=20.5$, is displayed. Each point $(t,y_{\rm
min})$ in Fig.~\ref{ymin} tells that at time $t$ the orbit attains a
new minimal value $y_{\rm min}$.

On the $(t,y_{\rm min})$ curve in Fig.~\ref{ymin}, an orbit makes
some steep ``steps'' and flat ``plateaus''. A steep step indicates
that the orbit diffuses very quickly in this stage, probably
crossing a well-developed chaotic region or jumping over a wide
island-chain. While a plateau indicates that the orbit spends a long
time achieving a small advance in $y$ direction, i.e., it is facing
an obstacle on its diffusion route. And the wider the plateau is,
the more effective the obstacle is (the stronger stickiness effect
it has).

Owning to the chaotic character, two orbits follow different
diffusion routes and may suffer different stickiness effects, that's
why in Fig.~\ref{ymin} two curves differ from each other. But in the
same phase space, the sticky regions that they have to pass through
are the same, thus the plateaus appear at the same $y_{\rm min}$ on
two curves. Note the logarithmic scale of the abscissa affects the
apparent widths of the plateaus.

When an orbit reaches a new minimum $y_{\rm min}$ at moment $t$
during its diffusion, a temporary (local) rotation number $\omega$
of the orbit around this point is calculated. For each $y_{\rm
min}$, we use 1001 orbital points (including the one right at moment
$t$, 500 before and 500 after $t$) to estimate $\omega$. These
records $(\omega, y_{\rm min})$ are plotted in Fig.~\ref{ymin} too,
as the curves crossing the panel from lower left to upper right. On
these curves, we can find the temporary rotation number ($\omega$,
indicated on the top $x$ axis) for each $y_{\rm min}$. Again, the
jumping of $\omega$ implies that the orbit is temporarily in the
vicinity of a big island, while the (nearly) continuous variation of
$\omega$ implies that the orbit is crossing a series of finely
ground island-chains, or a cantorus. The consistence between the
continuous variation of $\omega$ and the wide plateau of $y_{\rm
min}$ is evident in Fig.~\ref{ymin}.

Combining the explorations on the diffusion routes in \S~2.2 and in
Fig.~\ref{ymin}, we know that an orbit meets considerable resistance
when crosses some regions, where the $(t,y_{\rm min})$ curves make
plateaus and the $(\omega,y_{\rm min})$ curves vary continuously. We
choose three such ``sticky regions'' to investigate in detail in the
following part of this paper. Three horizontal lines in
Fig.~\ref{ymin} show the lower bounds of these regions.
Particularly, the lowest one is close to wide plateaus on the
$(t,y_{\rm min})$ curves, indicating a strong stickiness effect
around it. Checking the rotation number, we see that it is close to
the golden rate. To explain clearly the ``sticky region'', here we
define the sticky region labeled Region {\textbf c} as an example. A
lower bound $y_c^l=18.68$, as shown in Fig.~\ref{ymin}, is set
first. For an orbit starting from an initial point above this lower
bound, if the $y$ value of this orbit gets smaller than $y_c^l$ for
the first time during its diffusion, we regard the orbit as having
crossed Region {\textbf c}. An upper bound $y_c^u$ is selected as
well. When the $y$ value of an orbit gets smaller than $y_c^u$ for
the first time, the orbit is defined as having entered Region
{\textbf c}. The difference between the upper and lower bound
$\delta y_c = y_c^u - y_c^l$ is the width of this sticky region.

It is worth to note that there is no clear/sharp boundary for a
sticky region in the above definition. Although we use a $y_{\rm
min}$ value to denote it, we should keep in mind that a sticky
region is not a straight (but most probably curved) belt across the
phase space.

\subsection{Characteristic angle of hyperbolic structure and diffusion speed}

\begin{figure}[H]
 \vspace{10.0 cm}
 \includegraphics{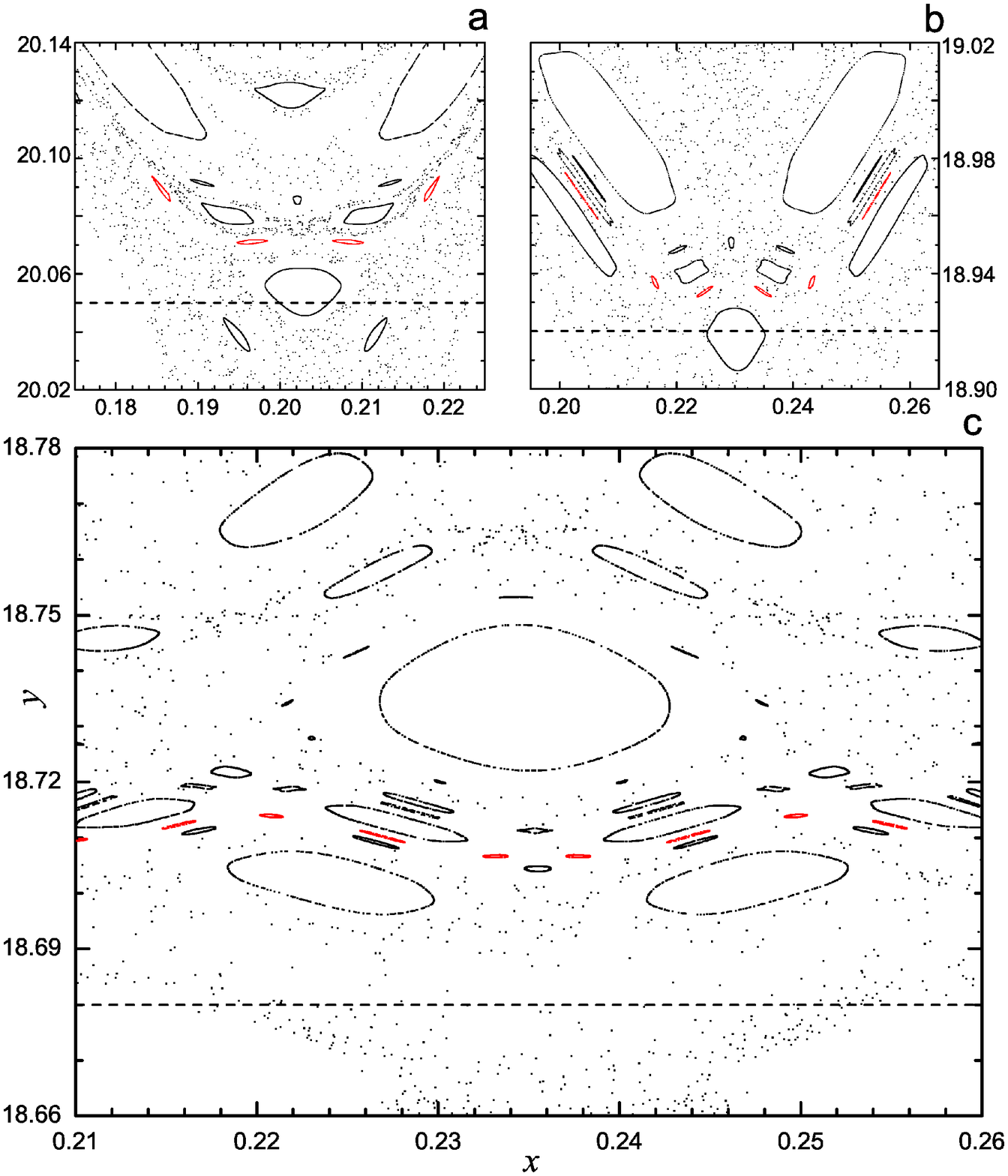}
 \caption{Three sticky regions. Three panels are for the regions indicated
 by arrows in Fig.~\ref{ymin} (see text). Horizontal lines in each panel represent
 the line of $y=20.05, y=18.92$ and $y=18.68$, respectively. The island chains in red
 are the ones with the highest period number found in the region. }
 \label{islchn}
\end{figure}

To explore the stickiness effects in these sticky regions, first we
plot in Fig.~\ref{islchn} part of the phase space in detail around
the corresponding local minimum $y_{\rm min}$ where orbits feel the
considerable resistance along their diffusion routes.

According to Birkhoff theorem [see e.g. 31], the elliptic periodic
orbits and hyperbolic periodic orbits emerge in pairs after the
breaking of the invariant tori in a 2D mapping. The elliptic
periodic orbits locate in the centers of islands, while the
hyperbolic periodic orbits exist between every two adjacent islands
in an island-chain. These two types of periodic orbits share the
same rotation number. The diffusion of an orbit is through these
hyperbolic structures. % and the stickiness effect is caused by them,
%as we show in our previous papers \citep{sun05,sun09}.

The elliptic periodic orbits are surrounded by invariant curves and
they together make islands. The hyperbolic periodic orbits however
are unstable and the geometric structures around them are not as
visible as the islands. So we plot in Fig.~\ref{islchn} as many as
possible the island chains to imply the positions of hyperbolic
periodic orbits and to show the composition of the sticky regions.
In each region, the island chain with the highest periods (they are
71, 73 and 144 in Region {\textbf a}, {\textbf b} and {\textbf c})
have been plotted in red. Of course, there must be islands of even
higher periods, but their sizes are much smaller, and it is
impossible to plot all of them in practice. The hyperbolic periodic
points can be located and the characteristic angles of the
hyperbolic structures can be calculated, using the algorithm
introduced in above section. Some characteristic angles in these
regions are summarized in Table~\ref{ang3}.

%---Begin of the Table ----------

\begin{table*}[t]
\caption{Characteristic angles of hyperbolic structures in three
sticky regions. The hyperbolic periodic orbits are denoted by their
rotation numbers, given in normal fraction numbers and continued
fractions. The angles are in degrees. } \center{
 \begin{tabular}{|l|l|l|}
 \multicolumn{3}{c}{Region {\textbf a}}  \\
 \hline
   \multicolumn{2}{|c|} {Rotation number}  & \hfill{} \\
 \cline{1-2}
 Fraction & Continued Fraction &  \raisebox{1.6ex}[0pt]{Angle $(\alpha)$} \\
 \hline
 165/61 & [2;1,2,2,1,1,3]   & 0.718653 \\
 119/44 & [2;1,2,2,1,1,2]   & 0.845155 \\
 192/71 & [2;1,2,2,1,1,1,2] & 0.772372 \\
 73/27  & [2;1,2,2,1,2]     & 1.094884 \\
 \hline
  \multicolumn{3}{c}{Region {\textbf b}}  \\
 \hline
 171/65 & [2;1,1,1,2,2,3]   & 0.717274 \\
 121/46 & [2;1,1,1,2,2,2]   & 0.863938 \\
 192/73 & [2;1,1,1,2,2,1,2] & 0.777229 \\
 71/27  & [2;1,1,1,2,3]     & 1.179671 \\
 \hline
 \multicolumn{3}{c}{Region {\textbf c}}  \\
 \hline
 199/76  & [2;1,1,1,1,1,1,1,3]     & 0.506420 \\
 343/131 & [2;1,1,1,1,1,1,1,2,2]   & 0.472460 \\
 144/55  & [2;1,1,1,1,1,1,1,2]     & 0.582802 \\
 377/144 & [2;1,1,1,1,1,1,1,1,1,2] & 0.477654 \\
 233/89  & [2;1,1,1,1,1,1,1,1,2]   & 0.534822 \\
 89/34   & [2;1,1,1,1,1,1,2]       & 0.741975 \\
 \hline
 \end{tabular}
 }

 \label{ang3}
\end{table*}
%-----end of the Table -------

We have seen from Table~\ref{examp} that the characteristic angles
of hyperbolic structures with higher period numbers are smaller.
This trend can be seen again in each region in Table~\ref{ang3}. But
when considering the characteristic angles with nearly the same
period number in different regions, for example, the characteristic
angles corresponding to the rotation number 192/71 in Region {\bf
a}, 192/73 in {\textbf b} and 199/76 in {\textbf c}, we find that
the angle in Region {\textbf c} $(0.506^\circ)$ is significantly
smaller than the other two in Regions {\textbf a} $(0.772^\circ)$
and {\textbf b} $(0.777^\circ)$, while the latter two are nearly
equal to each other. This result is expectable since Region {\textbf
c} makes the widest plateau in Fig.~\ref{ymin} and the diffusion
time through this region is in the order of $10^8 \sim 10^9$, much
longer than that in Regions {\textbf a} and {\textbf b}.

\begin{figure}[H]
 \vspace{5. cm}
 \includegraphics{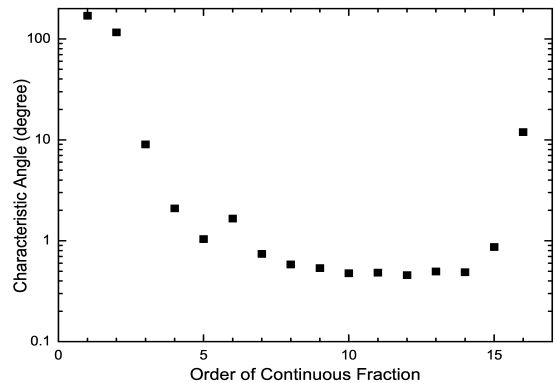}
 \caption{The lower bound of characteristic angles in Region {\textbf c}. The abscissa is the order
 of the truncated continued fraction, which is the rotation number of the hyperbolic
 periodic orbit. The ordinate is the corresponding characteristic angle.
 The last point of order 16 represents the hyperbolic periodic orbit with
 rotation number 10946/4181. }
 \label{convgm}
\end{figure}

The characteristic angle may decreases as the period of orbits
increases, but it is impossible in practice to compute the
characteristic angles for all the hyperbolic structures.
Fortunately, the information about the smallest characteristic angle
in a region can still be obtained. In Region {\textbf c}, the
rotation numbers of the hyperbolic periodic orbits is close to the
``golden rate'', and it is well-known that the KAM torus with the
golden rate rotation number is the most robust one. Meanwhile, it is
natural to assume that the ``smallest'' characteristic angle should
be found in the hyperbolic structures of the new-born cantorus just
after the breaking of KAM torus. Since the golden rate can be
approached by continued fractions, the characteristic angles of
hyperbolic structures with rotation numbers of truncated continued
fractions may approach to the characteristic angle in the cantorus.
Following this idea, we continue the calculations listed in
Table~\ref{examp} to higher order of the continued fraction. The
variation of the characteristic angle with respect to the order of
the continued fraction is illustrated in Fig.~\ref{convgm}. Note, as
the order increases the corresponding periodic orbits tend to gather
around the golden rate cantori, and no later than the 7th order
(89/34) all the hyperbolic structures under consideration accumulate
in Region {\textbf c}.

As shown in Fig.~\ref{convgm}, the calculations are stopped at the
16th order where the rotation number is 10946/4181. We cease the
calculations for two reasons as follows. First, the calculations so
far have shown that the characteristic angle has a lower bound when
the truncated continued fractions approach the golden rate. Even
higher order does not decrease this lower bound. Second,
calculations for the high order continued fractions (therefore high
period number) are very expensive in computer time, and it is
difficult to control the error in this case because we have to deal
with a large number of linear equations (Eq.~(15)). So we adopt the
smallest angle in Fig.~\ref{convgm} at the 12th order,
$\alpha_c=0.457^\circ$, as the lower bound of the characteristic
angle in Region {\textbf c}.

For Regions {\textbf a} and {\textbf b}, there is no golden rate
rotation number. However we can follow a similar process as we have
done for Region {\textbf c} to calculate the lower bound of
characteristic angles. We expand the continued fractions listed in
Table~\ref{ang3} by inserting number `1' into them right before the
last number, then we have series of truncated continued fractions of
different orders. Adopting these continued fractions as the rotation
numbers, we calculate the characteristic angles of the corresponding
hyperbolic structures. In this way, we obtain the lower bound of the
characteristic angles $\alpha_a = 0.665^\circ$ in Region {\textbf a}
and $\alpha_b = 0.667^\circ$ in Region {\textbf b}.

Since $\alpha_a \approx \alpha_b > \alpha_c$, we may derive directly
from these values that the diffusion in Region {\textbf c} is slower
than that in Regions {\textbf a} and {\textbf b}. In other words,
the stickiness effect in Region {\textbf c} is more intensive than
that in Regions {\textbf a} and {\textbf b}, while Regions {\textbf
a} and {\textbf b} have nearly the same stickiness effects. To
verify this conclusion, we carry out some calculations of diffusion
speed below.

\begin{figure}[H]
 \vspace{4.0cm}
 \includegraphics{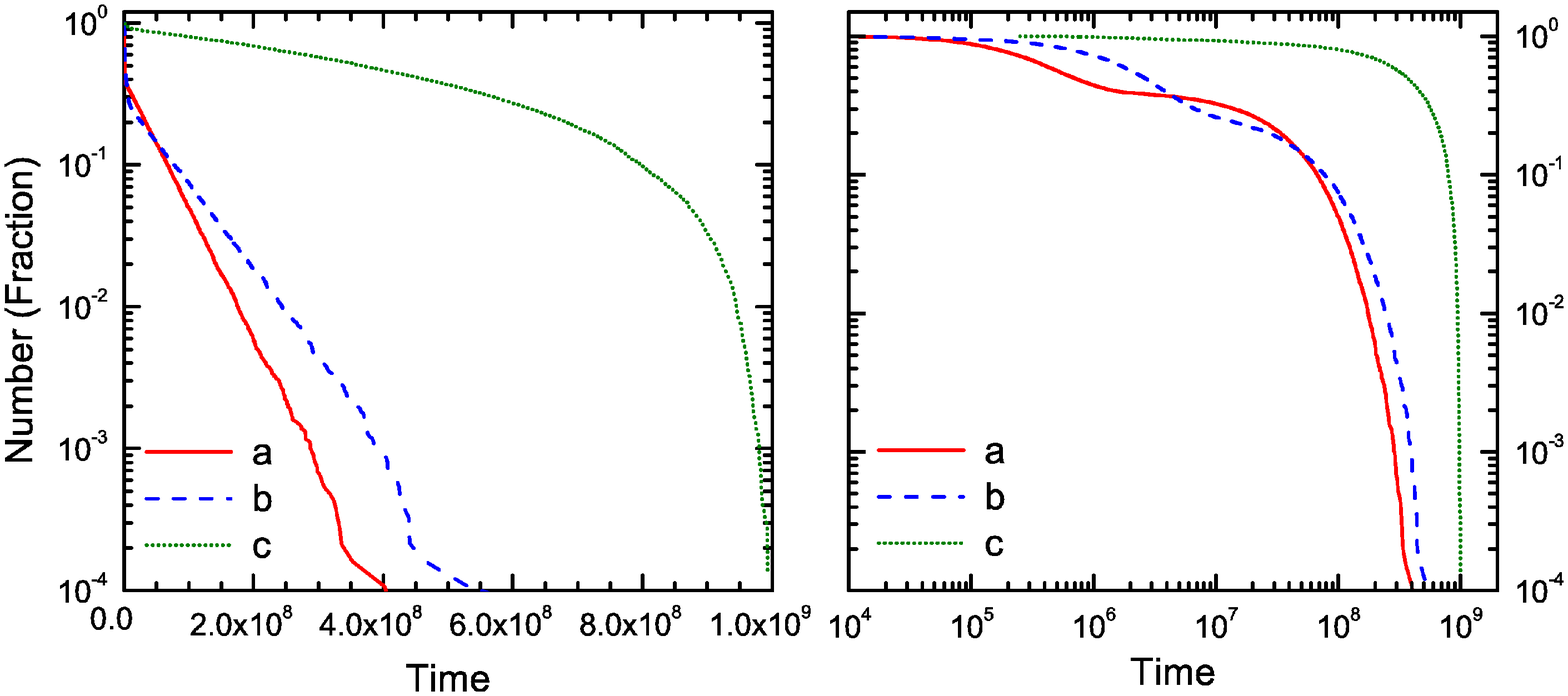}
 \caption{The diffusion speeds of orbits in Regions {\textbf a}, {\textbf b} and {\textbf c}.
 The time is given in linear scale in left panel, and logarithmic scale in right
 panel. }
 \label{esct}
\end{figure}

The diffusion speed, generally is measured by the decay rate of the
survival numbers of a bunch of orbits in a defined area [8, 16, 17].
We set randomly $40,000$ initial points in a rectangle from $(0.45,
26.5)$ to $(0.55,31.5)$ around the hyperbolic fixed point of
rotation number $3/1$ that is embedded in a developed chaotic area
and above the Regions {\textbf a}, {\textbf b} and {\textbf c} in
the phase space (see Fig.~\ref{minor}). All the orbits are iterated
for $10^9$ times using the mapping. The evolution of these forty
thousand orbits are then followed, especially, we monitor the
variation of the action variable $y$ of each orbit.

The definition of sticky region has been introduced in \S~4.1. For
Regions {\textbf a}, {\textbf b} and {\textbf c}, the lower
boundaries are chosen to be $y_a^l=20.05, y_b^l=18.92, y_c^l=18.68$
respectively, as indicated by the horizontal lines in
Figs.~\ref{ymin} and \ref{islchn}. After some tests, we set the
width of the sticky region $\delta y=0.1$ for all three regions. The
upper boundaries of the sticky regions are then defined as
$y_{a,b,c}^u= y_{a,b,c}^l + \delta y$.

An orbit may wander for quite a while before it enters a specific
area. To exclude the evolution history before an orbit entering the
vicinity of Region {\textbf a}, {\textbf b} or {\textbf c}, we
define the moment $t_a^u$ (or $t_b^u, t_c^u$) when for the first
time the $y$ value of an orbit is smaller than the upper boundary
$y_a^u$ (or $y_b^u, y_c^u$) as the moment of entering Region
{\textbf a} (or Region {\textbf b}, {\textbf c}). On the other hand,
when an orbit satisfies for the first time $y< y_a^l$ (or $y_b^l,
y_c^l$) at moment $t_a^l$ (or $t_b^l, t_c^l$), it is regarded as
having escaped from the sticky region. The time duration $\Delta
t_{a,b,c} = t_{a,b,c}^l - t_{a,b,c}^u$ then is defined as the
surviving time of an orbit in the corresponding sticky region.
Following the evolution of the 40,000 orbits, we record the
surviving time of each orbit in three sticky regions. With these
records, we count how many orbits survive in a specific region at a
given surviving time. Some orbits may have been initialized on
invariant curves by chance, and a small number of orbits may wander
in the chaotic phase space but never reach a specific region. These
orbits are excluded from the final statistics. After removing these
orbits and normalizing the number of orbits at the beginning to
unit, the diffusions in the three regions are summarized in
Fig.~\ref{esct}.

Apparently, the diffusion is quick in Regions {\textbf a} and {\bf
b}, but slow in Region {\textbf c}, reflecting that the stickiness
effect in Region {\textbf c} is much stronger than in other regions.
On the other hand, although Regions {\textbf a} and {\textbf b}
occupy different areas in the phase space and they comprise
different components as shown in Fig.~\ref{islchn}, they have nearly
the same diffusion speed.
%(thus bear the same stickiness effect).
The results are consistent with the conclusion we derived from the
characteristic angle values.
%In the left panel of Fig.~\ref{esct}
%the escaping curves for Region {\textbf a} and {\textbf b} seem
%different from each other, but in fact this difference appears only
%after tens of millions iterations and only about one tenth of orbits
%survive till then.

\subsection{Characteristic angle and perturbation parameter}

So far we have seen that in the phase space of a mapping with given
perturbation parameter $k$, the stickiness effects of different
sticky regions are related to the characteristic angles of
hyperbolic structures embedded in the regions. Smaller
characteristic angles lead to stronger stickiness effects and slower
diffusions. %The property of the mapping is controlled by the
%parameter ($k$ in Eq.~(6)), thus the characteristic angles change with $k$.
Below we will investigate the relation between the stickiness effect
and the characteristic angle when the parameter $k$ changes. We will
focus on the Region {\textbf c} where the diffusing orbits suffer
the most significant stickiness effects as shown above.

\begin{figure}[H]
 \vspace{5.5 cm}
 \includegraphics{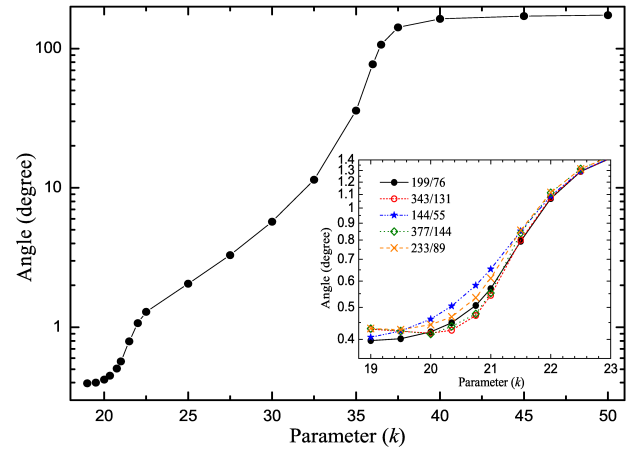}
 \caption{The variation of characteristic angle with
 respect to the parameter $k$. The solid circles show the angle of
 the hyperbolic structure with rotation number $199/76$. The embedded panel
 shows the characteristic angles of
 several hyperbolic structures with different rotation numbers. }
 \label{fang-k}
\end{figure}

In Region {\textbf c}, we select five hyperbolic periodic orbits to
analyze, including the one with rotation number 377/144 for which
the corresponding island-chain was plotted in red in
Fig.~\ref{islchn}, three above it with rotation numbers 199/76,
343/131 and 144/55, and one below it with the rotation number
233/89. The corresponding island chains are all plotted in
Fig.~\ref{islchn}, and they stay close to each other in a narrow
area in the phase space. Their characteristic angles at different
$k$ values are then calculated . The results are illustrated in
Fig.~\ref{fang-k} and some of them are listed in Table~\ref{tang-k}
too.

\begin{table*}[t]
\caption{The characteristic angles at different parameter $k$. The
angles are in degrees. Only five hyperbolic structures in Region
{\textbf c} are listed here. }
 \center{\begin{tabular}{|c|c|c|c|c|c|c|c|}
 \hline
   & \multicolumn{7}{c}{Angle $(\alpha)$} \vline \\
 \cline{2-8}
 \raisebox{1.6ex}[0pt]{Rot. Num.} & $k=19.00$ & $20.35$ & $21.50$ & $22.50$ & $27.50$ & $32.50$ & $36.00$ \\
 \hline
 199/76  & 0.396155 & 0.448865 & 0.792791 & 1.291957 & 3.288836 & 11.42482 & 77.06509 \\
 343/131 & 0.429506 & 0.426083 & 0.799331 & 1.294550 & 3.288836 & 11.42482 & 77.06509 \\
 144/55  & 0.405655 & 0.504247 & 0.851525 & 1.295668 & 3.288836 & 11.42482 & 77.06509 \\
 377/144 & 0.431027 & 0.437425 & 0.840257 & 1.314751 & 3.288844 & 11.42482 & 77.06509 \\
 233/89  & 0.431263 & 0.467751 & 0.856823 & 1.314805 & 3.288844 & 11.42482 & 77.06509 \\
 \hline
 \end{tabular}}

 \label{tang-k}
\end{table*}

When $k$ is small, the global KAM tori exist in Region {\textbf c}
and these hyperbolic structures are bounded by these tori.
%Consequently, the stable and unstable manifolds of the hyperbolic structures are ``narrowed'' to
%the parallel directions and the characteristic angles are small.
As pointed out in several literatures [e.g. 27, 36], in the vicinity
of an invariant torus in a 2D phase space, the resonances
(island-chains and hyperbolic periodic orbits) accumulate
geometrically toward the torus with their increasing orders (period
numbers). Owning to continuity, the directions of the stable and
unstable manifolds of the hyperbolic structures are ``forced'' to be
nearly parallel to the torus, thus the characteristic angles are
quite small in the close neighborhood of a torus. Here we see from
the embedded panel in Fig.~\ref{fang-k} that the angles seem to
approach a limit slowly as $k$ decreases. Consequently, orbits
starting from the close vicinity of an invariant torus will diffuse
along the resonance line (high-order island-chain) for a long time
before its leaving for adjoining resonances. As $k$ increases, the
chaos is developed and the characteristic angle increases.
Particularly, after the global KAM tori disappear in this region
(when $k\gtrsim 20.75$) the characteristic angle increases
dramatically. As a result, the diffusion speed is expected to be
enhanced dramatically.
% Finally, the characteristic angle approaches in the opposite direction to the upper limit ($180^\circ$).

Another interesting phenomenon is the difference between the
characteristic angles of different rotation numbers. When $k$ is
small, the characteristic angles are small and the relative
difference between different rotation numbers is considerable. But
the relative difference becomes smaller as the angles increase and
almost invisible when $\alpha \sim 2^\circ (k=25)$. This is clearly
illustrated in the embedded picture in Fig.~\ref{fang-k}. The reason
is that the difference of hyperbolicity between different periodic
orbits is apparent when $k$ is small, but it becomes ignorable when
$k$ is large. In the latter case, chaos has developed, making the
whole region uniformly hyperbolic. Thanks to this observation, to
show the angle variation in a wide $k$ range, it is enough to follow
just one of the hyperbolic structures, e.g., the one shown in
Fig.~\ref{fang-k} with rotation number $199/76$.

\subsection{Stickiness effect and characteristic angle}

The stickiness effect (or the diffusion speed) changes when the
characteristic angle changes with $k$. We will present the relation
between the diffusion speed and the size of characteristic angle in
this part. To show this correspondence, we need to define the
diffusion speed at first. The time needed by a bunch of orbits to
cross through a definite sticky region may be a good measure. But a
``sticky region'' may deform as the structure of phase space
distorts with the varying parameter $k$. Thus we check first the
variation of the position of a hyperbolic periodic orbit at
different $k$.

\begin{figure}[H]
 \vspace{5.5 cm}
 \includegraphics{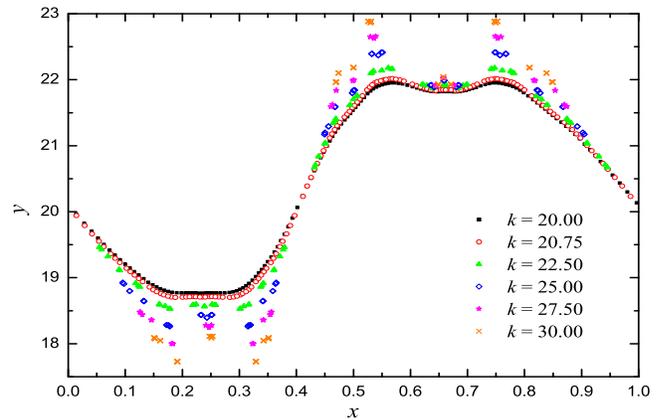}
 \caption{The positions of the hyperbolic periodic orbits of rotation number
 $377/144$. Cases of different $k$ are represented by different symbols. }
 \label{fpos-k}
\end{figure}

We plot in Fig.~\ref{fpos-k} the hyperbolic periodic orbit of
rotation number $377/144$ at six $k$ values. As $k$ increases, the
profile of the hyperbolic periodic orbit is distorted. The other
four hyperbolic periodic orbits of rotation numbers 199/76, 343/131,
144/55 and 233/89 are calculated too, and we find that their
profiles are distorted in a similar way as shown in
Fig.~\ref{fpos-k} and they always stay together within a narrow area
in the phase space. Therefore, the ``sticky region'' for different
$k$ can be defined in a similar way as in \S~4.2.

Just as we have done for the case $k=20.75$, forty thousand points
initialized in the rectangle area are followed again for different
$k$ values. We now focus only on the Region {\textbf c} where the
above mentioned five hyperbolic periodic orbits are embedded.
Knowing the locations of these hyperbolic structures (as shown in
Fig.~\ref{fpos-k}), the ``sticky region'' at different $k$ is
defined again. And also, the surviving time of an orbit in the
sticky region is given by the time duration $\Delta t = t^l - t^u$
as before.

\begin{figure}[H]
 \vspace{5.5 cm}
 \includegraphics{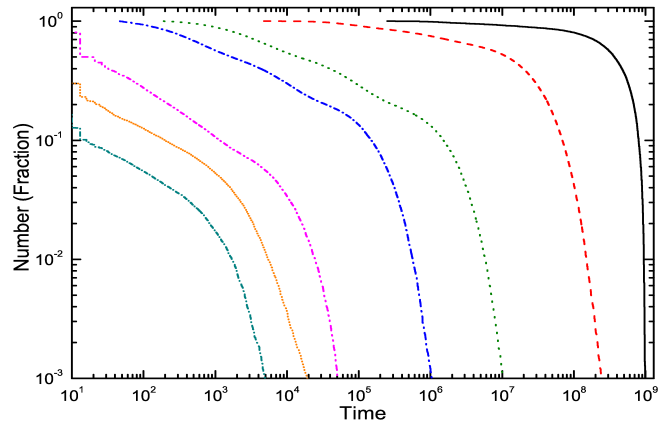}
 \caption{The diffusion at different $k$. From top to down (right to
 left), the curves represent the case for $k=20.75$, 21.00, 21.50, 22.50,
 25.00, 27.00 and $k=30.00$, respectively. }
 \label{fdif-k}
\end{figure}

All the orbits are iterated for $10^9$ times. We record the duration
time $\Delta t$ of each orbit, and it is regarded as the surviving
time of the orbit in the sticky region. The number of surviving
orbits changes with time. Following the same technique for plotting
Fig.~\ref{esct}b, we summarize our calculations for several $k$ in
Fig.~\ref{fdif-k}.

It is clear that the diffusion speeds up as $k$ increases. For the
case of $k=20.75$, the first orbit escaping happens at $t \sim
2\times 10^5$, when more than 90 percent of orbits have escaped in
the case of $k=22.50$. Nearly all orbits of $k=25.0$ and $k=27.00$
escape before $t\sim 10^4$, while only 1 percent of orbits survive
after only hundreds of iterations for $k=30.00$.

%We will show below how this diffusion speed is related to the
%characteristic angle of the hyperbolic structures involved.

In many literatures [e.g. 8, 16, 17, 35], the diffusion curves
similar to the ones in Fig.~\ref{fdif-k} were fitted by a power law
$N_{\rm survival}\propto t^{-z}$. The exponent $z$ then can serve as
a measurement of the diffusion speed (or the intensity of the
stickiness effect). But such a power law is not so distinct in our
cases as we see in Fig.~\ref{fdif-k}. Instead, we use directly the
time when most of the orbits have escaped to measure the diffusion
speed, (or inversely, the intensity of stickiness effect). More or
less arbitrarily, we adopt the moment when 90 percent of orbits have
escaped, denoted by $t_{1/10}$ since only one tenth of orbits
survive at this moment, as an estimation of the diffusion speed.
Thus a quick diffusion has a small $t_{1/10}$ while a large
$t_{1/10}$ indicates a slow diffusion (and strong stickiness
effect).

\begin{figure}[H]
 \vspace{5.5 cm}
 \includegraphics{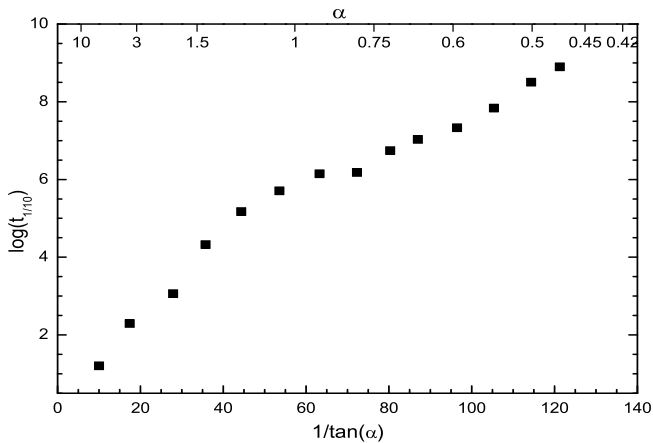}
 \caption{The variation of escaping time with respect to the characteristic angle
 ($\alpha$). The abscissa is $\cot\alpha=1/\tan{\alpha}$ and the ordinate is
 the escaping time ($t_{1/10}$, see text for definition) in logarithm. Several
 $\alpha$ values are indicated along the top axis too. }
 \label{fesc-a}
\end{figure}

For $k$ varying from 20.75 to 30.00, we calculate the $t_{1/10}$ and
the characteristic angles in this region at these $k$ values are
computed as well. They are illustrated in Fig.~\ref{fesc-a}. A small
characteristic angle leads to a slow diffusion, that is, a large
$t_{\rm 1/10}$. So, to visualize the variation of diffusion speed at
small characteristic angle, we use $1/\tan{\alpha}$ instead of
$\alpha$ itself as the abscissa in Fig.~\ref{fesc-a}, while several
$\alpha$ values are indicated along the top axis. The ordinate is
the diffusion speed $t_{1/10}$ in logarithm scale. For a given
parameter $k$, the characteristic angle in Fig.~\ref{fesc-a} is the
lower bound of the characteristic angles in Region {\textbf c}. They
are calculated in the same way that we have introduced in \S~4.2.
%using the truncated continued fractions as rotation numbers to approach the golden rate.

As indicated by Fig.~\ref{fesc-a}, the stickiness effect is
significant only when the characteristic angle is small. When the
characteristic angle $\alpha \sim 5^\circ (1/\tan\alpha = 11.5)$,
most of orbits have escaped just after tens of iterations, implying
a weak stickiness effect. On the other end of smaller $\alpha$ in
Fig.~\ref{fesc-a}, the escaping time will be longer than $10^9$ when
$\alpha$ is smaller than $0.45^\circ (1/\tan 0.45^\circ=127.3)$. An
orbit with such a long diffusion time can be practically regarded as
stable (never escape). In fact, the global KAM curves exist in
Region {\textbf c} when $k\lesssim 20.5$. And the characteristic
angle is smaller than $0.45^\circ$ when $k<20.5$, as we can see from
Fig.~\ref{fang-k}.

It is worth to note that the points in Fig.~\ref{fesc-a} may be
fitted by two linear functions with a crossover around $\alpha \sim
1^\circ$. The possible mechanism underneath such a relation deserves
an investigation in future.

When $k \gtrsim 20.75$, all the global KAM tori in Region {\textbf
c} have broken. The stickiness effect suffered by an orbit crossing
this region arises from the island-chains and cantori embedded in
the region. The diffusion speed is then determined by the
characteristic angle of hyperbolic structures among the
island-chains and cantori. Generally, in the vicinity of invariant
tori, hyperbolic structures accumulate, thus the stickiness effect
felt by orbits close to the tori, where the stickiness effect was
firstly found and defined \cite{kar82,kar83}, has the same origin:
the hyperbolic structures.

\section{Conclusions}

The stickiness effect in the phase space is an interesting
phenomenon with implications in many areas of sciences. It has
attracted many attention in the fields of mathematics, physics and
astronomy. Many structures such as the KAM tori, the island chains,
the cantori, and the hyperbolic structures in the phase space are
found to have stickiness effect. Among them, we think the hyperbolic
structures play the essential role, as we have shown in our previous
papers.

In this paper, we describe some geometric details of the hyperbolic
structure, especially the angle between the stable and unstable
manifolds, and relate these geometric characters with the strength
of the stickiness effect.

Usually the recursive algorithm is employed to calculate the
positions of periodic orbits in the phase space. But a usual
recursive algorithm fails when we try to find the hyperbolic
periodic orbits, because the set of hyperbolic periodic orbits is of
measure zero when the system is far from the integrable one. In this
paper, a numerical algorithm is introduced, with which we can
compute the precise locations of the hyperbolic periodic orbits. And
also, a practical algorithm for computing the directions of stable
and unstable manifolds of the hyperbolic structures is presented.

With these numerical algorithms, we have computed the angles between
the stable and unstable manifolds in different hyperbolic structures
in the 2D phase space of a mapping model. We investigate the
stickiness effect (diffusion speed) in different regions of the
phase space, and how it changes with the perturbation parameter.

Our findings in these calculations may be summarized as below:

- In an area where the last KAM torus has broken, the characteristic
angle between the stable and unstable manifolds of the hyperbolic
structure is related to the rotation number. The more irrational the
rotation number is, the smaller the angle is. But a lower bound of
this angle exists in a certain region.

- The stickiness effect (diffusion speed) is determined by the
characteristic angle. The smaller the angle is, the slower the
diffusion is, alternatively speaking, the stronger the stickiness
effect it has.

- A relationship between the characteristic angle of the hyperbolic
structure and the stickiness effect has been revealed. Nevertheless,
it is still difficult to find a quantitative way to describe this
relationship explicitly.

Although the above conclusions are drawn from the calculations in
regions consisted of island-chains and cantori but not global
invariant tori, they should be valid for all structures proven to
possess stickiness effect.

\Acknowledgements{\bahao This work was supported by the Natural
Science Foundation of China (NSFC, No.~10833001, No.~11073012) and
by Qing Lan Project (Jiangsu Province). Li and Sun are also
supported by the National Basic Research Program of China
(2013CB834103) and by NSFC (No.~11078001, No.~11003008)}.

%\vspace*{2mm}

\normalsize \vskip0.1in\parskip=0mm \baselineskip 18pt
\renewcommand{\baselinestretch}{1.06}\footnotesize\parindent=4mm\bahao

\vskip0.1in \noindent %{\normalsize \bf References}
\vskip0.1in\parskip=0mm

\REF{1\ }Karney C F, Rechester A B, White R B. Effect of noise on
the standard mapping. Physica D, 1982, 4: 425--438

\REF{2\ }Karney C F. Long-time correlations in the stochastic
regime. Physica D, 1983, 8: 360--380

\REF{3\ }Zaslavsky G M, Edelman M, Niyazov B A. Self-similarity,
renormalization, and phase space nonuniformity of Hamiltonian
chaotic dynamics. Chaos, 1997, 7: 159--181

\REF{4\ }Zaslavsky G M. Chaos, fractional kinetics, and anomalous
transport. Phys Rep, 2002, 371: 461--580

\REF{5\ }Barash O, Dana I. Type specification of stability islands
and chaotic stickiness. Phys Rev E, 2005, 71: 036222

\REF{6\ }Abdullaev S S. Chaotic transport in Hamiltonian systems
perturbed by a weak turbulent wave field. Phys Rev E, 2011, 84:
026204

\REF{7\ }Cust\'odio M S, Beims M W. Intrinsic stickiness and chaos
in open integrable billiards: Tiny border effects. Phys Rev E, 2011,
83: 056201

\REF{8\ }Jaff\'e C, Ross S D, Lo M W, et al. Statistical Theory of
Asteroid Escape Rates. Phys Rev Lett, 2002, 89: 011101

\REF{9\ }Karne C F, Bers A.  Stochastic ion heating by a
perpendicularly propagating electrostatic wave. Phys Rev Lett, 1977,
39: 550--554

\REF{10\ }Beloshapkin V V, Chernikov A A, Natenzon M Ia, et al.
Chaotic streamlines in pre-turbulent states. Nature 1989, 337:
133--137

\REF{11\ }Chia P-K, Schmitz L, Conn R W. Stochastic ion behaviour in
subharmonic and superharmonic electrostatic waves. Phys Plasmas,
1996, 3: 1545--1568

\REF{12\ }Fromhold T M, Patane A, Bujkiewicz S, et al. Chaotic
electron diffusion through stochastic webs enhances current flow in
super lattices. Nature, 2004, 428: 726--730

\REF{13\ }Varga I, Pollner P, Eckhardt B. Quantum Localization near
Bifurcations in Classically Chaotic Systems. Annalen der Physik,
1999, 8: 265-268

\REF{14\ }Kalapotharakos C, Voglis N, Contopoulos G. Chaos and
secular evolution of triaxial N-body galactic models due to an
imposed central mass. Astron Astrophys, 2004, 428: 905--923

\REF{15\ }Chirikov B, Shepelyansky D L. Correlation properties of
dynamical chaos in Hamiltonian systems. Physica D, 1984, 13:
395--400

\REF{16\ }Meiss J D, Ott E. Markov-tree model of intrinsic transport
in Hamiltonian systems. Phys Rev Lett, 1985, 55: 2741--2744

\REF{17\ }Lai Y C, Ding M C, Grebogi C, et al. Algebraic decay and
fluctuations of the decay exponent in Hamiltonian systems. Phys Rev
A, 1992, 46: 4661--4669

\REF{18\ }Perry A D, Wiggins S. KAM tori are very sticky: rigorous
lower bounds on the time to move away from an invariant Lagrangian
torus with linear flow. Physica D, 1994, 71: 102--121

\REF{19\ }Contopoulos G, Voglis N, Efthymiopoulos C, et al.
Transition spectra of dynamical systems. Celest Mech Dyn Astron,
1997, 67: 293--317

\REF{20\ }Froeschl\'e Cl, Lega E. Modeling mappings: an aim and a
tool for the study of dynamical systems. In: Analysis and Modeling
of Discrete Dynamical Systems. Benest D, Froeschl\'e Cl. eds. 1998,
3-54

\REF{21\ }Sun Y S, Fu Y N. Diffusion character in four-dimensional
volume preserving map. Celest Mech Dyn Astron, 1999, 73: 249--258

\REF{22\ }Contopoulos G, Harsoula M, Voglis N, et al. Destruction of
islands of stability. J Phys A: Math Gen, 1999, 32: 5213--5232

\REF{23\ }Contopoulos M, Harsoula M. Stickiness effects in chaos.
Celest Mech Dyn Astron, 2010, 107: 77--92

\REF{24\ }Efthymiopoulos C, Contopoulos G, Voglis N. Cantori,
islands and asymptotic curves in the stickiness region. Celest Mech
Dyn Astron, 1999, 73: 221--230

\REF{25\ }Sun Y S, Zhou L Y, Zhou J L. The role of hyperbolic
invariant sets in stickiness effects. Celest Mech Dyn Astron, 2005,
92: 257--272

\REF{26\ }Sun Y S, Zhou L Y. Stickiness in three-dimensional volume
preserving mappings. Celest Mech Dyn Astron, 2009, 103: 119--131

\REF{27\ }Morbidelli A, GiorgilliA. Superexponential stability of
KAM tori. J Stat Phys, 1995, 78: 1607--1617

\REF{28\ }Howard J E, Humpherys J. Nonmonotonic twist maps. Physica
D, 1995, 80: 256--276

\REF{29\ }Chirikov R V. A universal instability of many-dimensional
oscillator systems. Phys Rep, 1979, 52: 263--379

\REF{30\ }Cheng J, Sun Y S. Stable and Unstable Properties of the
Standard-liked Map. Random \& Computational Dynamics, 1996, 4:
73--85

\REF{31\ }Siegel C L, Moser J K. Lectures on Celestial Mechanics,
1971, Springer-Verlag, Berlin

\REF{32\ }Bangert V. Mather sets for twist maps and geodesics on
tori. Dynamics Reported, 1988, 1: 1¨C56

\REF{33\ }Cheng J. Variational approach to homoclinic orbits in
twist maps and an application to billiard systems. Z angew Math
Phys, 2004, 55: 400--419

\REF{34\ }Greene J M. A method for determining a stochastic
transition. J Math Phys, 1979, 20: 1183--1201

\REF{35\ }Benegeroles R. Universality of Algebraic Laws in
Hamiltonian Systems. Phys Rev Lett, 2009, 102: 064101

\REF{36\ }Zhou J L, Zhou L Y, Sun Y S. Hyperbolic Structures and the
Stickiness Effect. Chin Phys Lett, 2002, 19: 1254--1256

\end{multicols}

\end{document}